\documentstyle[psfig]{mn} 
\newcommand{\HI}{H\,{\sc i}}
\newcommand{\HII}{H\,{\sc ii}}
\newcommand{\kms}{km~s$^{-1}$}
\def\gapp{\ifmmode\stackrel{>}{_{\sim}}\else$\stackrel{>}{_{\sim}}$\fi}
\begin{document}

\title[\HI\ line measurements of pulsars]
{\HI\ line measurements of pulsars towards the Galactic Centre
and the electron density in the inner Galaxy}

\author[Johnston et al.]
{Simon Johnston$^1$, B\"arbel Koribalski$^2$, Joel M. Weisberg$^3$ \&
Warwick Wilson$^2$\\
$^1$QEII Fellow, School of Physics, University of Sydney,
NSW 2006, Australia\\
$^2$Australia Telescope National Facility, CSIRO, Epping, NSW 2121,
Australia\\
$^3$Carleton College, Dept. of Physics and Astronomy, Northfield, MN 55057, USA}

\maketitle

\begin{abstract}
We have measured 21-cm absorption and emission spectra in the direction of 
a further 7 southern pulsars with the Parkes telescope to derive 
their kinematic distances and to study the interstellar medium. 
For the first time we have successfully obtained \HI\ absorption measurements
for PSRs J1602--5100, J1740--3015 and J1745--3040. We have also 
significantly improved the sensitivity and resolution on PSRs J1600--5044,
J1752--2806, and J1825--0935, whose spectra have previously been measured,
and have corrected an error in the published distance to PSR J1824--1945.

Since the Frail \& Weisberg summary of pulsar distances in 1990, a 
further 23 pulsars now have measured \HI\ distances, mainly through the efforts
of the current group. We discuss the Taylor \& Cordes electron density 
model in light of these new measurements and find that, although
the model towards the Galactic Centre appears good, the line of
sight through the Carina spiral arm is poorly fit by the model.
\end{abstract}

\begin{keywords}
 pulsars: general --- radio lines: ISM --- stars: distances
\end{keywords}

\section{Introduction}
We obtained neutral hydrogen (\HI) emission and absorption spectra in the
direction of 7 southern pulsars to determine their
kinematic distances. This paper is the third in a series
on \HI\ observations of pulsars at the Parkes radio telescope.
We refer the reader to the earlier papers by
Koribalski et al. (1995; hereafter Paper\,I)
and Johnston et al. (1996; hereafter Paper\,II) \nocite{kjww95,jkww96}
for a comprehensive introduction as well as a full description
of the observations and data analysis.

The observations of \HI\ emission and absorption towards pulsars together with
a rotation curve or velocity field for the Galaxy allow the derivation of lower
and/or upper distance limits, depending on the distribution of \HI\ in the line
of sight. These distance limits can be coupled with the pulsar's
dispersion measure to obtain limits on the electron density along
the lines of sight. From knowledge of the electron density along many lines
of sight, a model can be constructed to convert dispersion measure into
distance with appropriate uncertainties.
The current model of Taylor \& Cordes (1993)\nocite{tc93}
represents the synthesis of
the observations to date and is widely accepted as the model of choice.
However, the data poorly constrain the model towards the Galactic Centre,
and the model has problems towards the Carina spiral arm
as discussed in Paper\,II.

The pulsars described here include three within 3\degr\ of the Galactic Centre
and the rest within 30\degr\ thereof. Combined with data from four
other pulsars \cite{wsfj95}, the current work allows a handle to 
be gained on the electron density in the inner Galaxy. As this is the
last paper in the series, we also discuss the 23 new \HI\ absorption
measurements since the advent of the Taylor \& Cordes (1993) model and
point out areas in which the model can be improved.

\section{Observations and data analysis}
\HI\ observations towards several southern pulsars were carried out 
on 1994 October 5--10 using the 64-m Parkes radio telescope.
The 20-cm receiver system, the correlator backend and the observing
procedure were all described in Paper\,I.
In brief, the digital correlator was configured into
2 $\times$ 512 channels (each with a bandwidth of 4 MHz) giving a velocity
resolution of 7.8 kHz or 1.74 \kms\ per channel. The data were reduced
off-line using standard procedures involving subtracting the `on-pulse'
spectra from the `off-pulse' spectra with suitable normalisation
(as detailed in Paper\,I).

Table~1 lists the pulsar parameters and details of the observing.
Columns 1 and 9 list the pulsar J2000 and B1950 names.
Columns 2 and 3 give the Galactic coordinates, columns 4 to 6 the pulsar
period, dispersion measure and flux density, respectively. Columns 7 and
8 give the integration time in minutes and the number of time
bins for each pulsar. In all cases the pulse occupied only 1 of these
phase bins (which was used to form the on-pulse spectrum) and the 
remaining were used for the formation of the off-pulse spectrum.

To calibrate the brightness temperature of the emission spectra
we used the Kerr et al. (1986)\nocite{kbjk86} emission survey
for PSRs J1600--5044 and J1602--5100, the Burton \& Liszt (1983)\nocite{bl83}
survey for PSRs J1740--3015, J1745--3040 and J1752--2806
and the Weaver \& Williams (1974a, 1974b)\nocite{ww74a,ww74b} survey for 
PSRs J1824--1945 and J1825--0935. To convert from the antenna temperatures
quoted in those papers to true brightness temperatures we used a 
correction factor of 1.025 for Kerr et al.,
1.52 for Burton \& Liszt and 1.14 for Weaver \& Williams.
Subject to the caveats contained in those
papers, we estimate the uncertainty in our brightness temperatures
to be of the order of 5\%. 

\section{Results on individual sources}
Figure~1 shows the emission and absorption spectra for each of the 7 pulsars.
Below the spectra we show the rotation curve appropriate
for the longitude and latitude of the pulsar.
In this paper we use an analytic expression given
by the best fit model of the Galactic rotation curve from Fich, Blitz,
\& Stark (1989)\nocite{fbs89}.
We adopt the standard IAU parameters \cite{klb86} for the distance
to the Galactic centre ($R_{0}$ = 8.5 kpc) and the solar orbital velocity 
($V_{0}$ = 220~\kms). The rotation
curve of the Galaxy is the azimuthally smoothed average of its velocity field
and thus implicitly assumes circular orbits around the Galactic centre.

The lower and upper distance limits of pulsars can be inferred from their
absorption and emission spectra, respectively. Following Frail \& Weisberg
(1990; Section V.a)\nocite{fw90}
we set the lower distance limit, $D_{\rm L}$, from
the centre of the farthest absorption feature in the spectrum and the upper
distance limit, $D_{\rm U}$, by the first emission peak above $T_{\rm B}$ =
35 K and beyond the most distant absorption feature.
The lower limit of 35 K on the brightness temperature is a conservative 
approach if one wishes to be certain of detecting absorption
\cite{rgmb72,wbr79}.
The simple Galactic rotation model cannot take
into account the presence of random or streaming motions in the \HI\ gas which
causes departures from purely circular rotation. One must allow for errors of
the order 7 \kms\ when converting the observed velocities into distances (see,
e.g., Weisberg et al. 1979; Shaver et al. 1982; Clifton et al. 1988).
\nocite{wbr79,sra+82,cfkw88}

For pulsars near the Galactic Centre, velocity crowding means that no
useful distances can be obtained directly from the Doppler shift of the
\HI\ line. However, comparison with other objects with known distance
and velocity can provide useful constraints as can the `peculiar'
velocity of the expanding 3 kpc arm. For all the objects discussed below
we use as much information from the literature
as possible when computing distance limits.

Table~2 shows the measured lower and upper velocity limits for each pulsar 
with the corresponding distance limits, where the distance errors are
computed assuming random motions of the \HI\ gas contribute 
7 \kms\ as described above. The electron density limits are then given
by the dispersion measure (Table~1) divided by the distance limits.
A discussion of the results for each pulsar is given below.

\subsection{PSR J1600--5044 (B1557--50); l,b=330\fdg7,1\fdg6}
Low-resolution \HI\ absorption spectra have been obtained by Ables \& 
Manchester (1976) and Manchester et al. (1981), but differ substantially.
The differences in optical depth near --40~\kms\ in the two
spectra are discussed in Deshpande et al. (1992) who claim this as evidence
for small-scale ISM structure. \nocite{am76,mwm81,dmra92}
The pulsar is close on the sky to PSR J1602--5100 discussed below.

Our measurements show, with much improved spectral resolution,
strong ($> 35$ K) \HI\ emission between --91~\kms\ and +30~\kms.
Several narrow absorption features are visible throughout this range
in addition to an astonishing absorption feature at --105~\kms.
It has an optical depth of 0.94 against an emission feature which
has a temperature of only 10 K! 
It is unusual to see such deep absorption in a relatively weak emission
feature at this longitude although similar absorption is sometimes
seen against the 3-kpc arm in the inner Galaxy.
The data yield a spin temperature of about 20 K, indicating that
the absorption cloud is unusually cold.
The \HI\ emission survey of Kerr et al. (1986) reveals nothing unusual in the
vicinity. The brightness temperature of the emission rises steeply towards 
the Galactic Plane at this velocity, peaking near 100 K.
One possiblity is that the absorption could be associated with the
small, cold, high density clouds which were recently detected in large
numbers in an infra-red galactic plane survey \cite{espc+98}.
However, these clouds are predominantly molecular \cite{ccep+98} and it
is unclear how much neutral atomic hydrogen would exist there.

The distance corresponding to this
feature is at 6.4$\pm$0.5 kpc which we assign as our lower distance
limit. This is somewhat more conservative than
Frail \& Weisberg (1990) \nocite{fw90} who assign the tangent
point distance of 7.4$\pm$1.1 kpc as the lower distance limit based
on critical analyses of the spectra of Ables \& Manchester (1976) and
Manchester et al. (1981).
At positive velocities, we might expect to see some absorption towards
the emission peak at +29.7~\kms, though its brightness is only 33 K.
From the lack of any significant absorption here we infer an 
upper distance limit of 18.2$\pm$1.2 kpc. Although the
emission temperature does not (quite) reach our 35 K cutoff
criterion, the relatively small noise in the absorption spectrum
strengthens our argument that the lack of observed absorption
supports the quoted upper limit.

Caswell et al. (1975)\nocite{cmr+75} observed \HI\ towards a 
number of (Galactic) sources between longitudes 
328\degr\ and 332\degr. All these are closer
to the Galactic plane than PSR J1600--5044 and in all cases emission
temperatures of $\sim$100 K are seen at --105 \kms.
The closest supernova remnant to the pulsar is SNR 330.2+1.0 \cite{chmw83}
but unfortunately it has not been observed in \HI\ nor has there been
any detection of maser lines associated with it \cite{gfgo97}.

\subsection{PSR J1602--5100 (B1558--50); l,b=330\fdg7,1\fdg3}
Frail \& Weisberg (1990) discuss an unpublished low-resolution spectrum 
for this pulsar by Manchester. They note the apparent absorption seen
near the tangent point velocity of --110~\kms. At this velocity the emission
temperature is only 12 K. However, assuming the absorption is real they
assign the tangent point distance of 7.4$\pm$1.1 kpc as a {\it lower} limit.

The emission spectrum is very similar to that of PSR J1600--5044 (described
above) with strong \HI\ emission between --90 and +30 \kms.
Our spectrum shows absorption at least up to --72~\kms\ corresponding
to a distance of 4.5$\pm$0.4 kpc.
There is no absorption seen against the reasonably 
bright feature at --90~\kms. Gas at this velocity can either be on the
near or far side of the tangent point, hence we set the upper limit to
be beyond the tangent point at a distance of 9.4$\pm$0.4 kpc.
However, we also have a hint of absorption near --110~\kms\ as in the
Manchester spectrum. This feature has an optical depth of 0.36 (4.6-$\sigma$).
Given that we see strong absorption at --105~\kms\ in PSR J1600-5044
(only 20 arcmin distant) against a similarly low temperature cloud we
assign a lower distance limit at the tangent point (7.4$\pm$0.5 kpc).

There are 2 known \HII\ regions located near the pulsar and significantly
off the Galactic plane. G331.354+1.072 has a recombination line
velocity of --79~\kms\ and G331.360+0.507 has a velocity
of --46~\kms\ in both recombination lines and H$_{\rm 2}$CO \cite{ch87b},
but no independent distances are known.

\subsection{PSR J1740--3015 (B1737--30); l,b=358\fdg3,0\fdg2 \\ 
~~~~~~PSR J1745--3040 (B1742--30); l,b=358\fdg6,--1\fdg0}
These pulsars are only 1\degr\ apart on the sky and lie within 2\degr\
of the Galactic centre. Along this line of sight,
velocity crowding is too severe to convert velocity to distance.
However, the presence of the 3 kpc arm at a velocity of $-60$~\kms\ 
(see Burton \& Liszt 1983\nocite{bl83}) helps to constrain the distance.

Both pulsars show similar emission profiles. Strong \HI\ emission
extends from --20~\kms\ to +20~\kms and there is a dip in the
brightness temperature due to self-absorption from a local cloud
\cite{cl84} at a velocity of +8~\kms.
Towards both pulsars, emission is seen from the 3 kpc arm at --60~\kms\
although the brightness temperature is significantly less than 30 K in 
the case of PSR J1745--3040 and only $\sim$30 K for PSR J1740--3015.

The absorption profiles of the pulsars differ in that absorption with
an optical depth of $\sim$0.5 is seen at --15~\kms\ towards 
PSR J1740--3015 but not towards PSR J1745--3040.
No absorption is seen against the 3 kpc arm in either pulsar.
Dickey et al. (1983)\nocite{dkgh83} and Garwood \& Dickey (1989)\nocite{gd89}
show \HI\ spectra towards 3 background sources (1739-298, 1741-312
and 356.905+0.082) all less than 2\degr\ from the pulsars.
Absorption at --15~\kms\ and against the 3 kpc arm is clearly seen.
SNR G357.7--0.1 (MSH 17--39; the Tornado) shows similar absorption
characteristics - deep absorption features from +20 to --15~\kms\ and 
absorption at the location of the 3 kpc arm at --60~\kms\ \cite{rgmb72}.
This remnant is located beyond the 3 kpc arm, and possibly beyond
the Galactic Centre \cite{fgrg+96}. The other complex of interest in
this region of sky contains the SNR G359.1--0.5, the `Mouse' (G359.2--0.8)
and the `Snake' (G359.1--0.2). \HI\ observations of this whole area
by Uchida et al. (1992)\nocite{umy92}, show that the Snake and the SNR
lie beyond the 3 kpc arm but the Mouse shows no absorption against
the gas in the arm and is presumed to be a foreground object.

Given the evidence above, we should certainly see absorption against 
PSR J1740--3015 if it were located behind the 3 kpc arm, and we do not. 
Following Weisberg et al. (1995)\nocite{wsfj95},
therefore, we assign the 3 kpc arm at 5.5$\pm$0.6 kpc
as the upper distance limit. Given that absorption is seen against
the emission at --15~\kms\ we conclude that this feature lies in front
of the 3 kpc arm but at an unknown distance.
As PSR J1745--3040 does not show absorption
at --15~\kms\ in spite of a high emission temperature we conclude
that it must be in front of the arm, in front of PSR J1740--3015 and
also in front of the --15~\kms\ feature. Unfortunately, the lack
of any further kinematic information in this region forces us to assign
an upper distance limit of 5.5$\pm$0.6 kpc for this pulsar also.

The upper limits derived are consistent with the dispersion measure
distances to the pulsars and we note also that PSR J1745--3040
has a lower dispersion measure than PSR J1740--3015 as expected
if it is closer.

Saravanan et al. (1996) also observed both of these pulsars.  Their absorption
spectrum on PSR J1740--3015 is similar to ours, although ours has less noise
in spite of a slightly shorter integration time. We also do not see
any absorption against the weak emission feature at --180 \kms\ that
is visible in their figure nor would any absorption be expected
at this velocity given that the pulsar is closer than the 3-kpc arm.
(Saravanan et al. do not set any distance limits on the basis of 
this feature.)
For PSR J1745--3040, our integration time was longer than theirs and
our spectrum less noisy. Although the absorption spectra appear different,
given the relatively high noise levels in the spectra they appear
to be consistent at the 3-$\sigma$ level.

\subsection{PSR J1752--2806 (B1749--28); l,b=1\fdg5,--1\fdg0}
This pulsar was one of the first discovered \cite{tv68}.
A low resolution \HI\ absorption spectrum by
Guelin et al. (1969)\nocite{gghw69} did not clearly show absorption,
but a Gordon et al. (1969)\nocite{ggs69} spectrum with a resolution
of 3.8 \kms\ shows a single absorption feature at +9 \kms,
coincident with the self-absorption feature in the emission spectrum. 

This pulsar scintillates rather strongly through the 4 MHz band
and, in contrast to the other pulsars in this sample, we fitted
the absorption profile baseline (excluding the channels of interest)
with a 7th order polynomial to remove the effects of the scintillation.
Our data show three clear absorption features at velocities of
+4.9, +8.2 and +14.8 \kms. The feature at +8.2 \kms\ is exactly coincident
with the self-absorption feature in the emission profile and is the
same as that seen by Gordon et al. (1969). We note that if our spectrum
is smoothed to the same velocity resolution as that of Gordon et al.,
the three lines blend together into a single feature.
The features at both +5 and +9 \kms\ are due to a cold cloud at a
distance of 125 pc \cite{cl84} and Frail \& Weisberg (1990)
adopted this as the lower limit to the pulsar's distance.

We note, however, several other objects in this region with similar
velocities. The \HII\ region RCW 142 (G0.55--0.85) has a recombination line 
velocity of +13.3 \kms, an OH maser at +16.8 \kms\ and H$_{\rm 2}$CO velocities
of +15.6 and +20.8 \kms\ \cite{gw75}. Two nearby reflection nebulae 
(RCW 137 and 141) have H$\alpha$ velocities of +12.2 and +9.6 \kms,
respectively \cite{gg70b}. Finally, the OH/IR star IRAS 17482--2824
(Galactic coordinates 1.0, --0.83) has a maser velocity of +10 \kms\ and 
a photometrically derived distance of 1.2 kpc \cite{loe95}.
It is appealing to claim that because we see absorption beyond +10 \kms\
in our spectrum we could raise the lower distance limit to the
pulsar to 1.2 kpc.  However, the peculiar motion of the OH/IR star is
uncertain and the large spread in velocities of the features in RCW 142
forces us to adopt a cautious approach and hence we leave the lower
distance limit as 0.125 kpc. If distances were measured to any of the
RCW regions listed above the distance limit could be significantly improved.

\subsection{PSR J1824--1945 (B1821--19); l,b=12\fdg3,--3\fdg1}
Saravanan et al. (1996)\nocite{sdw+96} observed this pulsar in \HI\ and derived
a lower distance limit of 2 kpc. In our spectrum
\HI\ in emission extends from +5 to +35~\kms.
\HI\ absorption is seen over the same velocity range
and the last absorption peak at +26.4~\kms\ defines the
lower distance limit with 3.2$\pm$0.5 kpc. The lack of \HI\ above
30 K beyond this point implies no upper distance limit can be set.
We note that the 4-$\sigma$ absorption against the emission feature
at $+35$~\kms\ is marginally significant. However, the low brightness
temperature of this feature (25 K) and the lack of absorption at this
velocity in the Saravanan et al. (1996) spectrum makes it unlikely
that the absorption is real.
The conversion from velocity to distance is significantly in error in 
Saravanan et al. (1996) and our result defines the new distance limit.

An OH/IR star IRAS 18176--1848 lies about 1\degr\ away from the pulsar,
has a maser velocity of 13 \kms\ and a photometrically derived
distance of 2.6 kpc \cite{loe95} consistent with the above result,
if we also allow for a 7 \kms\ component of random velocity in the maser.
Gathier et al. (1986)\nocite{gpg86} display 21--cm absorption profiles of 
the planetary nebula NGC 6578 at galactic coordinates 10\fdg8,--1\fdg8 and
another source at 11\fdg3,--1\fdg6. However, since their absorption
cuts off at lower velocities than does the pulsar's,
they provide no additional useful  kinematic information.

\subsection{PSR J1825--0935 (B1822--09); l,b=21\fdg5,1\fdg3}
G\'omez-Gonz\`ales \& Gu\'elin (1974)\nocite{gg74} obtained a 
low-resolution ($\Delta v\sim 13$ \kms) \HI\
spectrum towards this pulsar, from which they derive only an upper distance
limit of 1.5 kpc. In our high resolution spectrum we see
strong \HI\ emission between zero and +38~\kms. The absorption is
seen only against the gas at zero velocity and hence no lower
velocity limit can be set. There is no significant absorption
against either the feature at +20~\kms\ or +35~\kms.
Dickey et al. (1983)\nocite{dkgh83} observed two nearby (extragalactic) sources
within $\sim$1\degr\ of the pulsar but at slightly higher
galactic latitudes. Both 1817--098 (G20.7+2.3) and 1819--096 (G21.0+2.0)
show deep absorption ($\tau > 1$) out to +35~\kms.

Garwood \& Dickey (1989)\nocite{gd89},
Radhakrishnan et al. (1972)\nocite{rgmb72} and
Caswell et al. (1975)\nocite{cmr+75} show \HI\ observations to 4 other sources
(G19.6--0.2, G21.5--0.9, G24.0+0.2 and G24.8+0.1)
within $\sim$2\degr\ of the pulsar.
Again, deep absorption is seen out to at least +60~\kms\ in all the sources,
although their lower galactic latitude indicates that they are not
the ideal comparison sources that the first two are.

We thus assign
an upper distance limit of 1.9$\pm$0.4 kpc based on the lack of
absorption against the emission feature at +20~\kms.

\section{Discussion}
The review of Frail \& Weisberg (1990)\nocite{fw90}
contained 50 pulsars for which \HI\ lower and/or
upper distance limits had been measured. Shortly thereafter,
Taylor \& Cordes (1993)\nocite{tc93}
developed a model of the electron density distribution in the Galaxy
to allow for the conversion from a pulsar's dispersion measure to its
distance. Since then, 28 pulsars have been observed in \HI\ of which
18 were observed for the first time and 5 had significantly different
\HI\ kinematic distances to those derived in older literature
(Frail et al. 1993\nocite{fkv93}, Paper I,
Weisberg et al. 1995, Paper II, Saravanan et al. 1996, this work).
The pulsars mainly lie in the 4th quadrant of the Galaxy. The
tangent points of three major spiral arms lie in this quadrant
(in order of increasing distance from the Sun and decreasing distance from
the Galactic Centre; the Carina, Crux, and Norma arms), but 
previous \HI\ observations of pulsars were few and far between.
The most recent discussion and summary of new results can be found in
Weisberg (1996).\nocite{wei96}

We can extend the work of Weisberg et al. (1995) in order to see
how the Taylor \& Cordes (1993)\nocite{tc93} model matches up to the
new results. Somewhat surprisingly, the distance derived by the model
to 11 of the 24 pulsars lie outside the measured \HI\ distances, 7 
on the low side and 4 on the high side.
Figure 2 displays this in graphical form. In the Figure,
the pulsars are projected onto an X-Y plane with the Galactic Centre at 0,0
and the Sun at 0,8.5. The location of the spiral arms included by
Taylor \& Cordes in their model are shown on the plot as is their
inner annulus of enhanced electron density.
Solid diamonds represent the Taylor \& Cordes (1993) dispersion
measure derived positions. The solid lines indicate the range of \HI\ measured
kinematic positions; those with an arrow have no upper distance limit.
The two pathological cases with very large DM-derived distances compared
to the \HI\ distance, PSRs J1056--6258 and J1801--2305,
are almost certainly located behind \HII\ regions which contribute 
substantially to the dispersion measure
(Paper II; Frail et al. 1993).

It is interesting to note that for most of the Galactic Centre pulsars,
the model does a reasonable job. However, very few pulsars lie beyond 
the expanding arm in this direction and even fewer lie on the 
far side of the Galactic Centre. Taylor \& Cordes (1993) have made the
point that this does not allow them to distinguish whether the 
electron density reaches a peak in an annulus at some radius from 
the Galactic Centre or continues increasing until the Centre.

Of the 8 pulsars between longitudes 300\degr\ and
342\degr\ (along the tangents to the Sagittarius-Carina and Scutum-Crux arms),
no fewer than 6 are not well fitted by the Taylor \&
Cordes model and all 6 have \HI\ distances {\it larger} 
than those calculated from the model.
It is evident from the figure that the Taylor \& Cordes model tends 
to assign pulsars to spiral arms because of the large electron density 
excess in those regions in the model,
so that a relatively large range of dispersion measure corresponds
to a small change in distance. In particular the model seems to overestimate
the electron density in the Sagittarius-Carina arm.

A further aid to determining distances to pulsars located behind spiral
arms is to compare their scintillation speeds with their
proper motions \cite{gup95}. The scintillation speed has a different
distance dependence to the proper motion velocity and also depends
on the location of the scattering screen (in this case presumably the spiral
arm). Unfortunately not many pulsars have both scintillation and proper
motion measurements. One example is PSR J1453--6413. Its dispersion
measure distance of 1.8 kpc is at odds with the \HI\ lower limit of
2.5 kpc. If we assume the spiral arm is 1.5 kpc from Earth in this
direction, then to match the scintillation speed \cite{jnk98} with
the proper motion \cite{bmk+90b} requires a pulsar distance of 3.3 kpc,
consistent with the \HI\ lower limit. Although this is just one example
(but see also McClure-Griffiths et al. 1995\nocite{mjsn98}),
it may be that towards the very clumpy regions of the spiral arms
that scintillation parameters are more appropriate than dispersion
measure for assigning distances to those pulsars
for which \HI\ absorption measurements are impractical. Ideally,
one requires \HI, scintillation and proper motion measurements of a large
number of pulsars in order to significantly refine 
further the electron density model of the Galaxy.

\section{Conclusions}
This paper completes our analysis of \HI\ observations towards 22
southern hemisphere pulsars started in Papers\,I and II.
The main results from these three papers are:
\begin{enumerate}
\item The Taylor \& Cordes (1993) model does a good job in assigning
distances to pulsars behind the Gum Nebula.
\item As found in other studies, the \HI\ gas is seen at velocities
forbidden by the standard Galactic rotation model along directions
tangent to the Carina spiral arm.
\item PSR J1709--4429, one of the few pulsars observable at TeV energies,
has a well-defined \HI\ distance which appears
to rule out the association with a nearby supernova remnant.
\item At least on the near side of the expanding 3 kpc arm, the
Taylor \& Cordes distance model works well towards the Galactic Centre.
\item Dispersion measure is not necessarily a good indicator of distance
when looking through the spiral arms, and modifications to the distance model
appear to be necessary in the region $300\degr < l < 345\degr$.
\end{enumerate}

\section*{Acknowledgments}
JMW was supported by NSF Grant AST9530710.
The Australia Telescope is funded by the Commonwealth of Australia
for operation as a National Facility managed by the CSIRO.

\bibliographystyle{mn}
\bibliography{modrefs,psrrefs,crossrefs}
\clearpage
\newpage
\begin{table*} 
\caption[]{Pulsar parameters and observing information}
\begin{flushleft}
\begin{tabular}{crrcccccc}
\hline 
 PSR      & $l~$   & $b~$   & DM            & Period & S$_{\rm 20}$ & T     & bins &  PSR  \\
 (J2000 ) &        &        &(pc\,cm$^{-3}$)& (ms)   & (mJy)        & (min) &      & (B1950)\\
\hline
1600--5044 & 330\fdg7 &  1\fdg6 & 262.8 & 192.60 & 15 & 232 & 16 & 1557--50 \\
1602--5100 & 330\fdg7 &  1\fdg3 & 169.5 & 864.20 &  6 & 398 & 32 & 1558--50 \\
1740--3015 & 358\fdg3 &  0\fdg2 & 153.0 & 606.64 &  6 & 175 & 32 & 1737--30 \\
1745--3040 & 358\fdg6 &--1\fdg0 &  88.8 & 367.43 & 14 & 224 & 16 & 1742--30 \\
1752--2806 &   1\fdg5 &--1\fdg0 &  50.4 & 562.56 & 35 & 50  & 32 & 1749--28 \\
1824--1945 &  12\fdg3 &--3\fdg1 & 224.3 & 189.33 &  9 & 324 & 16 & 1821--19 \\ 
1825--0935 &  21\fdg5 &  1\fdg3 &  19.9 & 768.97 & 11 & 427 & 16 & 1822--09 \\ 
\hline
\end{tabular}
\end{flushleft}
\end{table*}
\begin{table*} 
\caption[]{\HI\ kinematic distances for 7 pulsars}
\begin{flushleft}
\begin{tabular}{cccccc}
\hline
  PSR     &$v_{\rm L}$&$v_{\rm U}$&$D_{\rm L}$ & $D_{\rm U}$ & $n_{\rm e}$ \\
 (J2000)  &  (\kms)   &  (\kms)  &   (kpc)     &    (kpc)    & (cm$^{-3}$) \\
\hline
1600--5044& --105  & +29.7  & 6.4$\pm$0.5 & 18.2$\pm$1.2 & 0.014 -- 0.041 \\
1602--5100& --110  & --90   & 7.4$\pm$0.5 &  9.4$\pm$0.4 & 0.018 -- 0.023 \\
1740--3015& --14.8 & --60   & ---         &  5.5$\pm$0.6 & $>$0.028       \\
1745--3040&  ---   & --20   & ---         &  5.5$\pm$0.6 & $>$0.016       \\
1752--2806&  +14.8 &  ---   & 0.125       &   ---        & $<$0.40        \\
1824--1945&  +26.4 &  ---   & 3.2$\pm$0.5 &   ---        & $<$0.070       \\
1825--0935&  ---   &  +19.8 & ---         &  1.9$\pm$0.4 & $>$0.010       \\
\hline
\end{tabular}
\end{flushleft}
\end{table*}
\begin{table*} 
\caption[]{New \HI\ distances since Frail \& Weisberg (1990)}
\begin{flushleft}
\begin{tabular}{ccccccc}
\hline
  PSR     & $l$     & $b$     & $D_{\rm L}$ & $D_{\rm U}$ & Ref. & PSR \\
 (J2000)  & (\degr) & (\degr) &   (kpc)     &    (kpc)    &      & (B1950) \\
\hline
0742--2822& 243\fdg8 &--2\fdg4 & 2.0$\pm$0.6 & 6.9$\pm$0.8  & 2 & 0740--28 \\
0907--5157& 272\fdg2 &--3\fdg0  & ---         & 8.0          & 5 & 0905--51 \\
0908--4913& 270\fdg3 &--1\fdg0 & 2.4$\pm$1.6 & 6.7$\pm$0.7  & 2 & 0906--49 \\
0942--5552& 278\fdg6 &--2\fdg2 & ---         & 7.5$\pm$0.7  & 4 & 0940--55 \\
1048--5832& 287\fdg4 &  0\fdg6 & 2.5$\pm$0.5 & 5.6$\pm$0.8  & 4 & 1046--58 \\
1056--6258& 290\fdg3 &--3\fdg0 & 2.5$\pm$0.5 & 2.9$\pm$0.5  & 2 & 1054--62 \\
1224--6407& 300\fdg0 &--1\fdg4 & 4.3$\pm$1.4 &11.4$\pm$0.7  & 4 & 1221--63 \\
1326--5859& 307\fdg5 &  3\fdg6 & 3.0         & ---          & 5 & 1323--58 \\
1401--6357& 310\fdg6 &--2\fdg1 & 1.6$\pm$0.5 & 2.7$\pm$0.7  & 4 & 1358--63 \\
1453--6413& 315\fdg7 &--4\fdg4 & 2.5$\pm$0.5 &  ---         & 2 & 1449--64 \\
1559--4438& 334\fdg5 &  6\fdg4 & 2.0$\pm$0.5 &  ---         & 2 & 1556--44 \\
1600--5044& 330\fdg7 &  1\fdg6 & 6.4$\pm$0.5 &18.2$\pm$1.2  & 6 & 1557--50 \\
1602--5100& 330\fdg7 &  1\fdg3 & 7.4$\pm$0.5 & 9.4$\pm$0.4  & 6 & 1558--50 \\
1651--4246& 342\fdg5 &  0\fdg9 & 4.8$\pm$0.3 & ---          & 3 & 1648--42 \\
1707--4053& 345\fdg7 &--2\fdg2 & 3.8$\pm$0.5 & ---          & 3 & 1703--40 \\
1709--4429& 343\fdg1 &--2\fdg7 & 2.4$\pm$0.6 & 3.2$\pm$0.4  & 2 & 1706--44 \\
1721--3532& 351\fdg7 &  0\fdg7 & 4.4$\pm$0.5 & 5.2$\pm$0.6  & 3 & 1718--35 \\ 
1740--3015& 358\fdg3 &  0\fdg2 & ---         & 5.5$\pm$0.6  & 6 & 1737--30 \\
1745--3040& 358\fdg6 &--1\fdg0 & ---         & 5.5$\pm$0.6  & 6 & 1742--30 \\
1801--2305&   6\fdg8 &--0\fdg1 & 3.5         & 6.9          & 1 & 1758--23 \\
1824--1945&  12\fdg3 &--3\fdg1 & 3.2$\pm$0.5 &  ---         & 6 & 1821--19 \\
1825--0935&  21\fdg5 &  1\fdg3 & ---         & 1.9$\pm$0.4  & 6 & 1822--09 \\
1833--0827&  23\fdg4 &  0\fdg1 & 4.0$\pm$0.4 & 5.3$\pm$0.3  & 3 & 1830--08 \\
\hline
\end{tabular}
\end{flushleft}

\noindent
References: 1 - Frail et al. (1993), 2 - Paper I, 3 - Weisberg et al. (1995),
            4 - Paper II, 5 - Saravanan et al. (1996), 6 - this paper.
\end{table*}
\clearpage
\newpage
\begin{figure*}
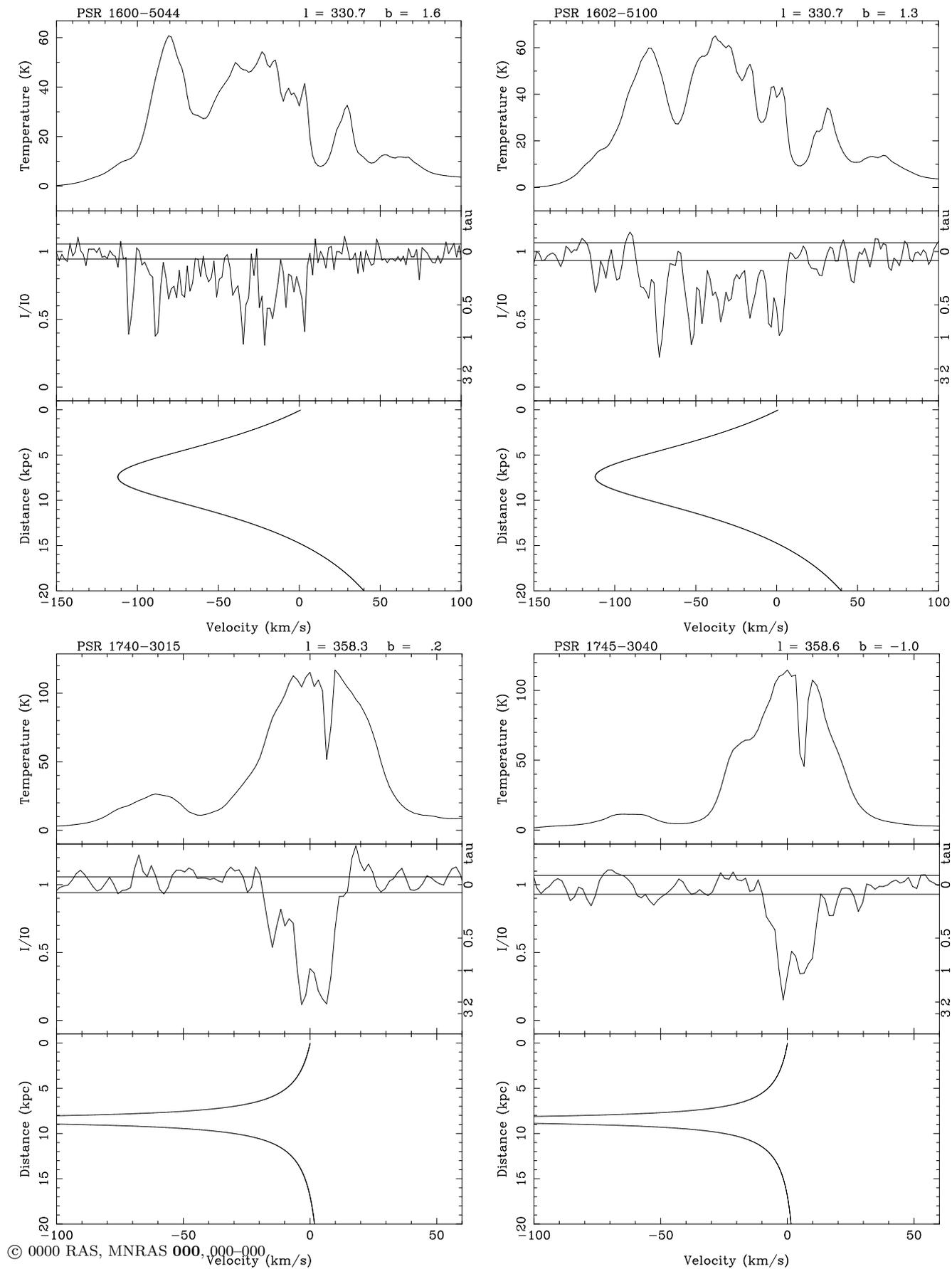
 
\caption{\HI\ spectra in the direction of seven southern pulsars.
	 The pulsar name (J2000) is given on top of each plot.
	 Each panel shows the \HI\ emission spectrum on top, 
	 the weighted \HI\ absorption spectrum in the middle, and the 
	 corresponding part of the Galaxy rotation curve on the bottom.
	 See the text for further description.}
%
\centering
\begin{tabular}{cc}
\psfig{figure=1600_final.ps,width=8.5cm}&\psfig{figure=1602_final.ps,width=8.5cm}\\
\psfig{figure=1740_final.ps,width=8.5cm}&\psfig{figure=1745_final.ps,width=8.5cm}\\
\end{tabular}
\end{figure*}
\clearpage
\newpage
\begin{figure}
\centering
\begin{tabular}{cc}
\psfig{figure=1752_final.ps,width=8.5cm}&\psfig{figure=1824_final.ps,width=8.5cm}\\
\psfig{figure=1825_final.ps,width=8.5cm}\\
\end{tabular}
\end{figure}
\clearpage
\newpage
\begin{figure*} 
\caption{Distance limits on 24 mostly inner Galaxy pulsars measured 
since the Taylor \& Cordes (1993) model was created.
The ends of each line represent the upper and lower limits along a 
particular line of sight, while the diamond shows the model-estimated distance.
An arrowhead on the end of a bar indicates that no upper distance limit
was determinable. The thick curves show the location of the
four spiral arms as delineated by Taylor \& Cordes; from top to bottom
the Perseus arm, the Sagittarius-Carina arm, the Scutum-Crux arm and the
Norma arm. We also show the inner annulus of enhanced electron density
derived in their model.
See Table 3 for details on the individual pulsars plotted here.}
\centering
\psfig{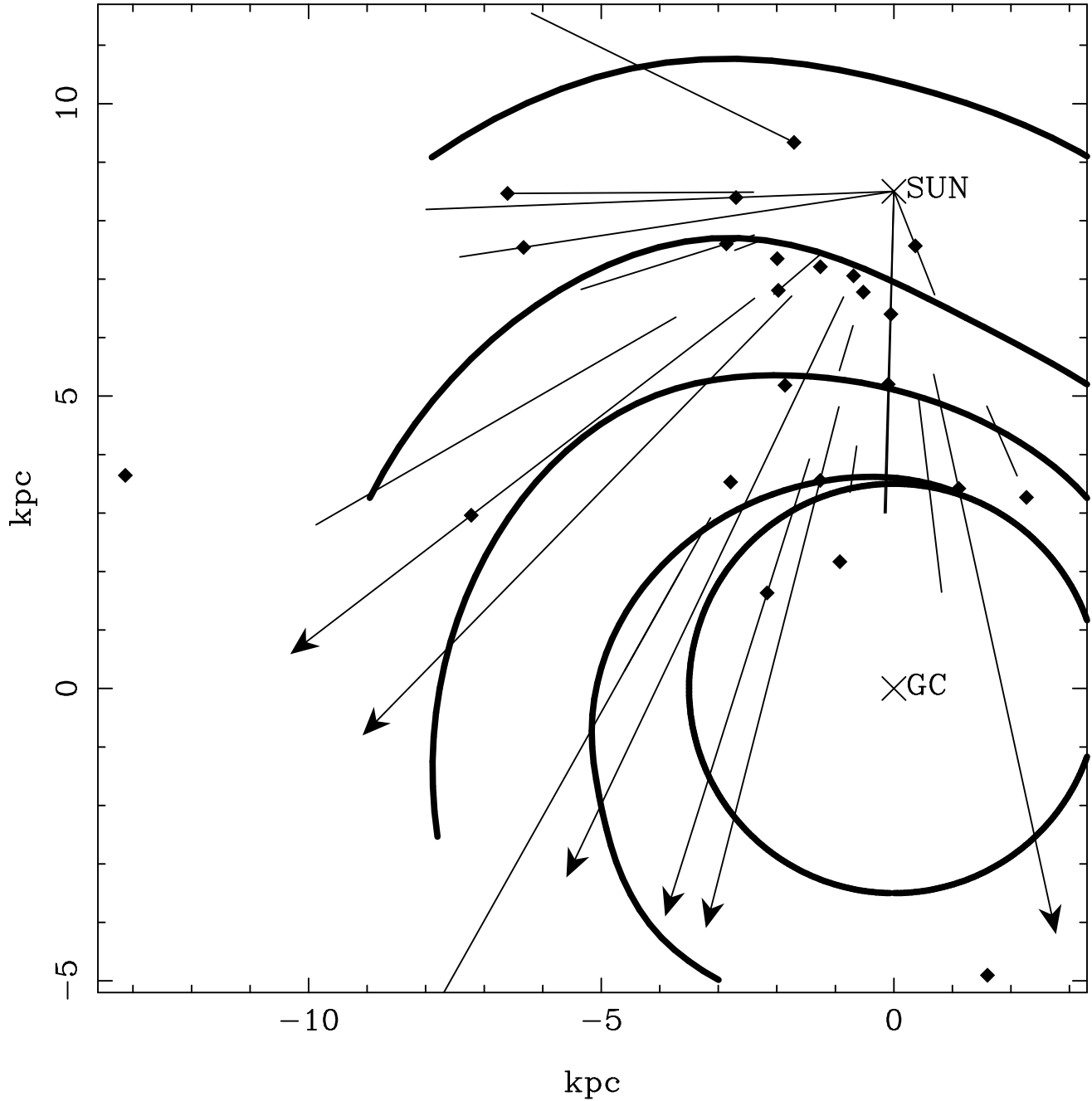}
\end{figure*}
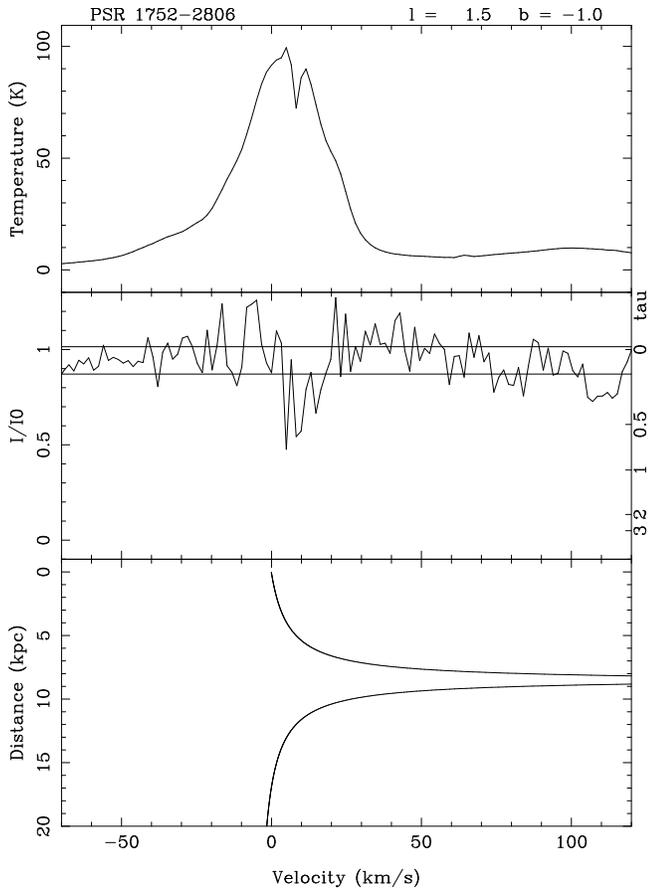
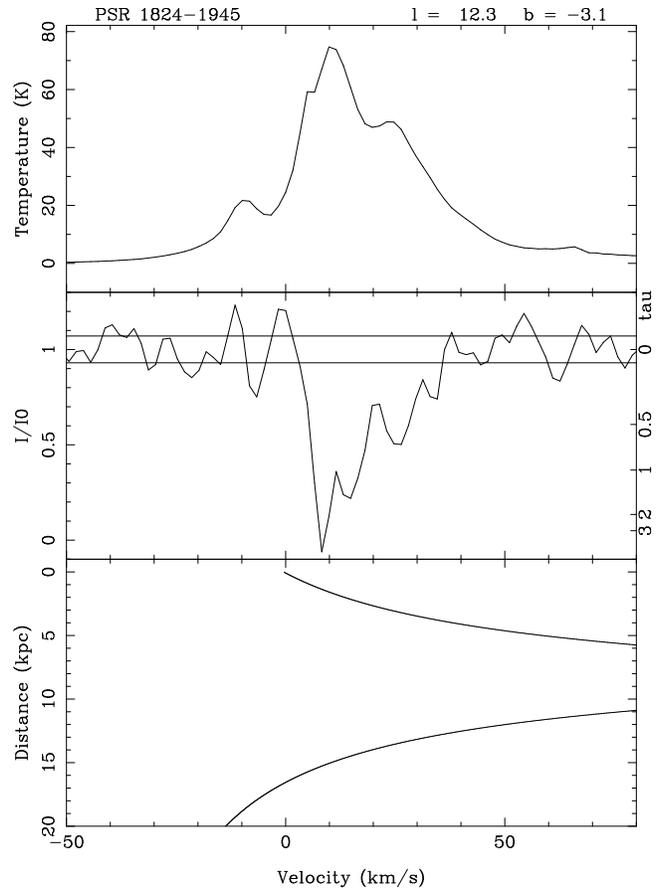
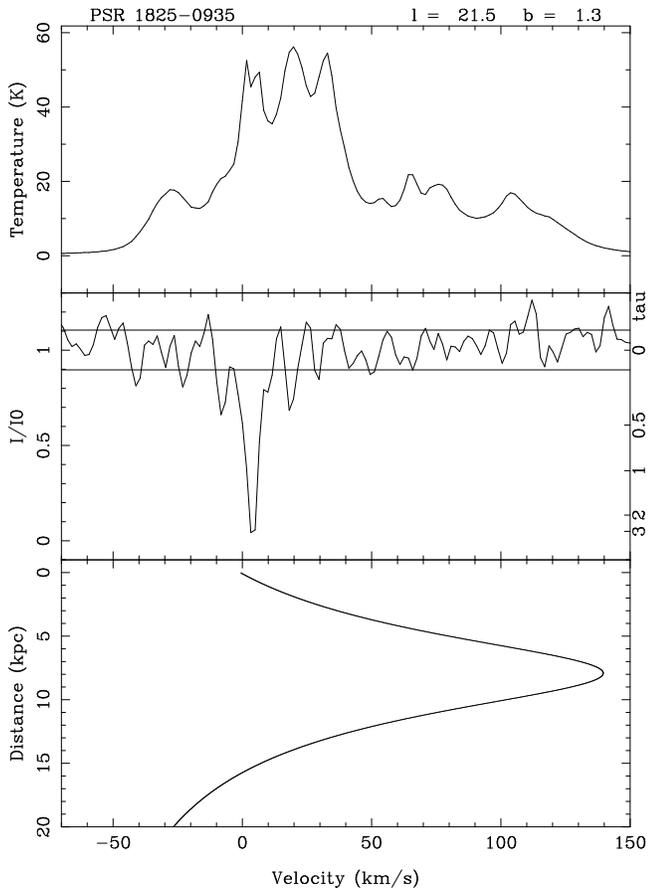
\end{document}